# Decreasing the Surgical Errors by Neurostimulation of Primary Motor Cortex and the Associated Brain Activation via Neuroimaging.

Yuanyuan Gao[1], Lora Cavuoto[2], Anirban Dutta[3], Uwe Kruger[1,4], Pingkun Yan[4], Arun Nemani[1,4], Jack E. Norfleet[5,6,7], Basiel A. Makled[5,6,7], Jessica Silvestri[2], Steven Schwaitzberg[2,8,9], Xavier Intes[1,4], and Suvranu De[1,4,*]

*1 Center for Modeling, Simulation and Imaging in Medicine, Rensselaer Polytechnic Institute, Troy, New York, USA*

*2 Jacobs School of Medicine and Biomedical Sciences, University at Buffalo, The State University of New York, Buffalo, New York, USA*

*3 Department of Biomedical Engineering, University at Buffalo, The State University of New York, Buffalo, New York, USA*

*4 Department of Biomedical Engineering, Rensselaer Polytechnic Institute, Troy, New York, USA*

*5 U.S. Army Combat Capabilities Development Command - Soldier Center (CCDC SC), Orlando, Florida, USA.*

*6 SFC Paul Ray Smith Simulation & Training Technology Center (STTC), Orlando, Florida, USA.*

*7 Medical Simulation Research Branch (MSRB), Orlando, Florida, USA.*

*8 Department of Surgery, University at Buffalo, The State University of New York, Buffalo, New York, USA*

*9 Buffalo General Hospital, Buffalo, New York, USA*

\* Corresponding author: Suvranu De.

**Email:** des@rpi.edu

**Keywords**

**Author Contributions**

Y.G., A.N., and S.D. designed research; Y.G., L.C., and J.S. performed research; Y.G. U.K., P.Y., and X.I. analyzed data; and Y.G. wrote the paper; and all the authors discussed the results and revised the paper.

**Abstract**

Acquisition of fine motor skills is a time-consuming process as it requires frequent repetitions. Transcranial electrical stimulation is a promising means of enhancing simple motor skill development via neuromodulatory mechanisms. Here, we report that non-invasive neurostimulation facilitates the learning of complex fine bimanual motor skills associated with a surgical task. During the training of 17 medical students on the Fundamentals of Laparoscopic Surgery (FLS) pattern cutting task over a period of 12 days, we observed that transcranial direct current stimulation (tDCS) decreased the error level and the variability in performance, compared to the Sham group. By concurrently monitoring the cortical activations of the subjects via functional near-infrared spectroscopy (fNIRS), our study showed that the cortical activation significantly stimulated by tDCS. The lowered performance error and the increased brain activation were retained after one-month post-training. This work supports the use of tDCS to enhance performance accuracy in fine bimanual motor tasks.

**Main Text**

**Introduction**

From learning to play the violin to performing delicate surgery, perfecting fine motor skills requires significant repetitive practice (1). The training may take days, months, and even years. Sometimes, despite repeated practice, the resulting skill level might remain low (2, 3). Deliberate practice, *i.e.*, purposeful practice that requires focused attention and feedback, has been proposed as a learner-centric approach to accelerate performance (4). However, the use of novel technology to enhance fine motor skills remains limited.

Recently, neuromodulation has been proposed to enhance motor skill learning. This is motivated by the finding that motor learning involves neuroplasticity (5). It is also shown that motor learning recruits multiple brain areas (6). Transcranial electrical stimulation (tES) is a neuromodulation technique that can affect neuroplasticity and can facilitate motor learning. It changes the excitability of the cortex by delivering a small amount of current using electrodes attached to the scalp. Studies have shown that tES improves human motor learning, including visuospatial learning, sequence learning, and adaptation, in its direct current form, transcranial direct current stimulation (tDCS). For example, during a five-day training program, the primary motor cortex (M1) region tDCS increased the performance scores for a visuospatial task (7). In another three-day training program, M1 tDCS increased performance scores for both sequence and visuospatial learning (8). More recently, transcranial random noise stimulation (tRNS) has shown to improve learning (20). The study in (20) confirmed that applying tRNS benefited the reaction time for a finger-tapping task. However, further studies are scarce.

Most tES studies have focused on simple unimanual motor sequence learning. The effect of tES on complex motor skills, such as bimanual motor skills, remains relatively under-studied. Although complex motor skills may be decomposed into simpler motor tasks (9), higher-level coordination is involved (10). Furthermore, the learning procedure for complex skills is usually time and resource-consuming. Here, to the best of our knowledge, we are the first to investigate the effect of tES on the learning procedure of a complex surgical motor task (11), which typically takes more than ten days to achieve proficiency (12).

Alongside the performance change in the motor learning procedure, we assessed the brain activation change considering its association with motor learning neuroplasticity. Prior fMRI studies (6) show that specific cortical areas are activated during the motor learning stages, including the prefrontal cortex (PFC), supplementary motor area (SMA), and M1 regions. To acquire the brain activation changes, we used a non-invasive functional brain imaging technique, functional near-infrared spectroscopy (fNIRS). This fNIRS technique has been widely used in other motor skill studies (13). In addition to high temporal- and spatial- resolution, fNIRS can be coupled with tES during motor tasks without constraining or interfering with motor task execution (14).

In our study, we tested the hypothesis that tES of the primary motor cortex will facilitate complex surgical motor skill learning. We report the behavioral metrics in initial learning, consolidation learning, and skill retention. The second hypothesis is that the tES changes brain activation. By testing the two hypotheses, we report the effects of M1 tDCS to be reducing the performance error, as well as stabilizing the trial-to-trial variability, in conjunction with increased brain activation in the left M1 region.

## Results

**Behavioral change in motor learning due to neuromodulation**

We collected the behavioral data from three groups of medical students (tDCS group: n = 5; tRNS group: n = 5; Sham group: n = 7) in their Fundamentals of Laparoscopic Surgery (FLS) pattern cutting task training procedure (see Methods and Fig. 1). We first present the results analyzing the performance error (Fig. 2). The tDCS group consistently reduced performance error (linear fit: p < 0.0001. Fig. 2c) and the performance error was significantly lower than the Sham group on day 4, and from day 7 to day 12 ($p < 0.05$) (Fig. 2a). The Sham group maintained a similar error level throughout the training (linear fit: p = 0.713. Fig. 2c). The overall error level of the tRNS group was significantly lower than Sham from day 2 (Fig. 2a), then no further improvement (linear fit: p = 0.667. Fig. 2c). The mean trial-to-trial variability is quantified by the mean of the standard deviation of the performance error across subjects on each day (Fig. 2b). Compared to the Sham group, the standard deviation values were lower in the stimulation groups (tDCS and tRNS) (significantly on day 8), indicating better ability to reduce the trial-to-trial variability.

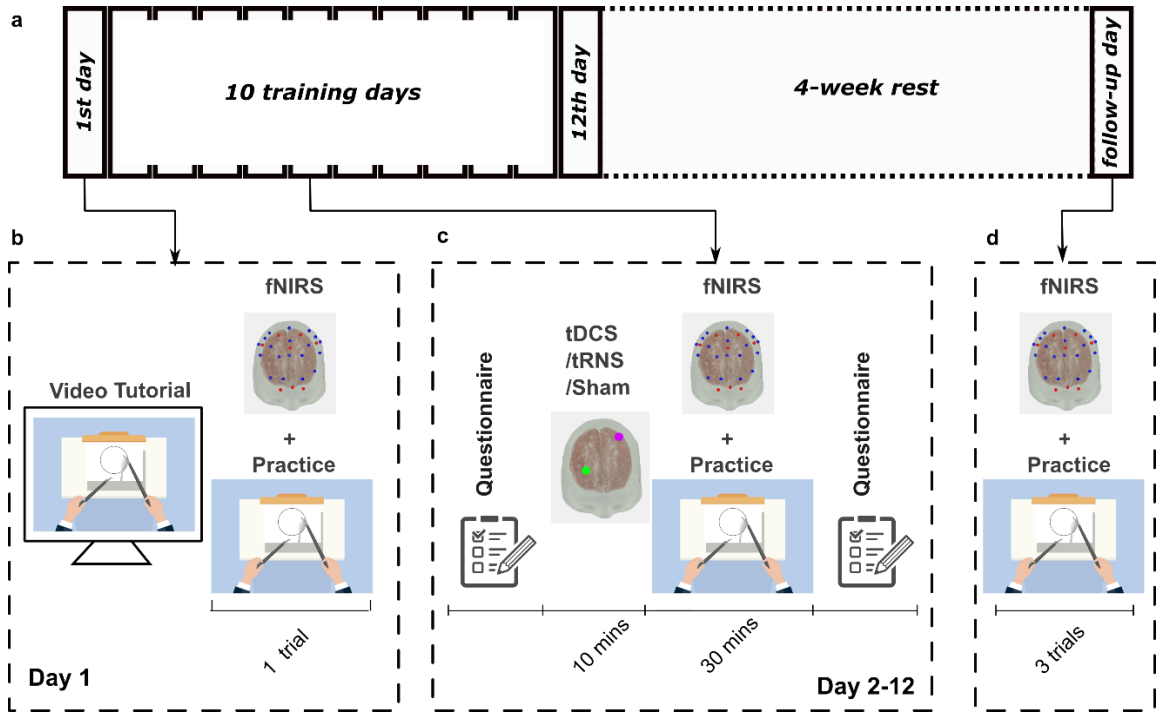

**Figure 1.** The experimental protocol designs. (a) Schematic showing the experimental design for this study. All participants went through a 12-day training procedure followed by a 4-week rest, and then attended a follow-up visit. (b) On the first training day, they watched a video tutorial on how to perform an FLS precision cutting task (https://youtu.be/mUBZoSO3KA8). Then they practiced one trial of the task on an FLS box trainer when the fNIRS data were acquired simultaneously. (c) From day 2 to day 12, on each day, they filled out a safety questionnaire and accepted 10 mins neuromodulation (tDCS/tRNS/Sham according to their assigned group) while they were resting. After the neuromodulation, they practiced the precision cutting task for 30 mins with fNIRS recording at the same time. At the end of the training session, they filled out the safety questionnaire again. (d) On the follow-up visit, to test their skill retention, participants performed the same FLS task for three trials.

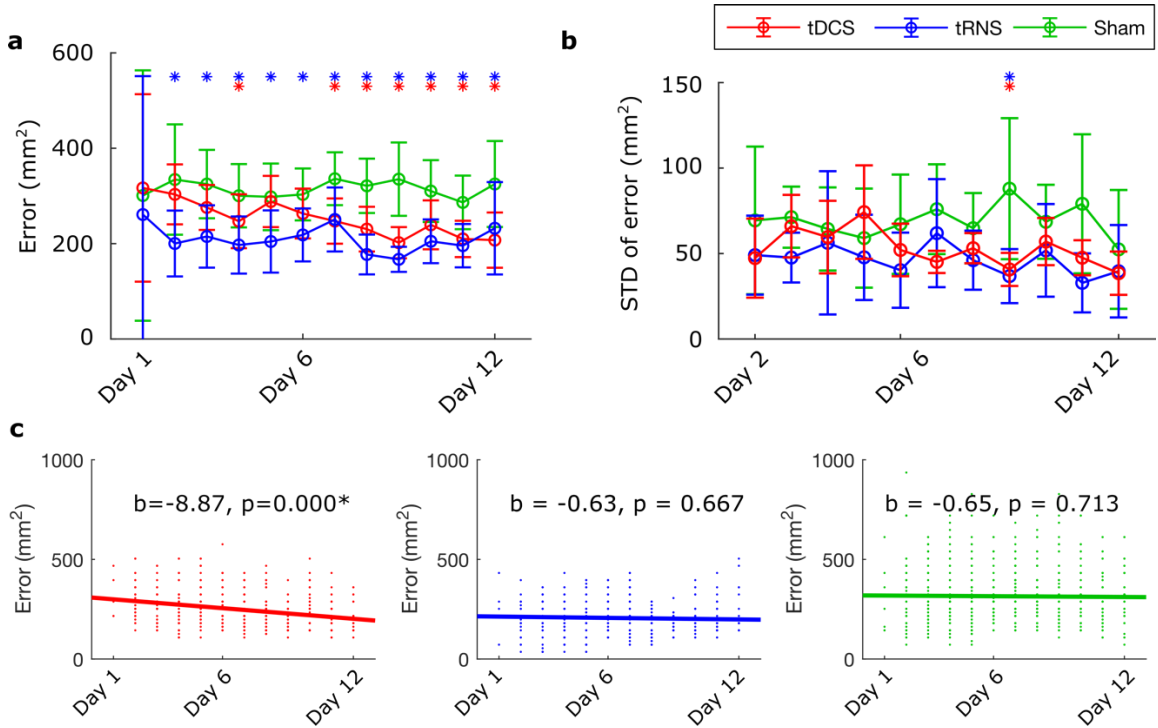

**Figure 2.** Bimanual motor task performance error. (a) The cutting error defined as the cutting area deviated from the pre-marked circle on the gauze for the 12 training days and the follow-up tasks are presented. (b) The mean trial-to-trial standard deviation (STD) value of error from day 2 to day 12. (c-e) The linear fitted line, the slot value b, and the significance p-values are presented for (c) tRNS, (d) tDCS, and (e) Sham. The red lines represent the tDCS group, the blue lines represent the tRNS group, and the green lines represent the Sham group. The stars represent a significant difference ($\alpha = 0.05$) compared to the Sham group.

We present the time and FLS score metrics (described in Methods, Motor task and task performance metrics) in Fig. 3. The three groups had similar learning curves in time and FLS scores (Fig. 3a, 3b). The tRNS group performed slower than the other two groups in the middle of the training (days 4, 7, and 8, in Fig. 3a). As expected, the longer performance times yielded lower FLS scores (day 7 and day 8, in Fig. 3b). Compared to the other two groups, the FLS scores in the tDCS group were higher during the initial learning stage but did not show a significant difference until day 5 (days 5, 6, and 9, Fig. 3b). At the end of the training, after day 9, the three groups did not show a significant difference in both performance time and FLS score. The standard deviations of the performance time, and FLS score, were also calculated on day 2-12 (Fig. 3c,3d). The tDCS group performed lowering trend of trial-to-trial standard deviation of the performance time and FLS score than the tRNS and Sham group.

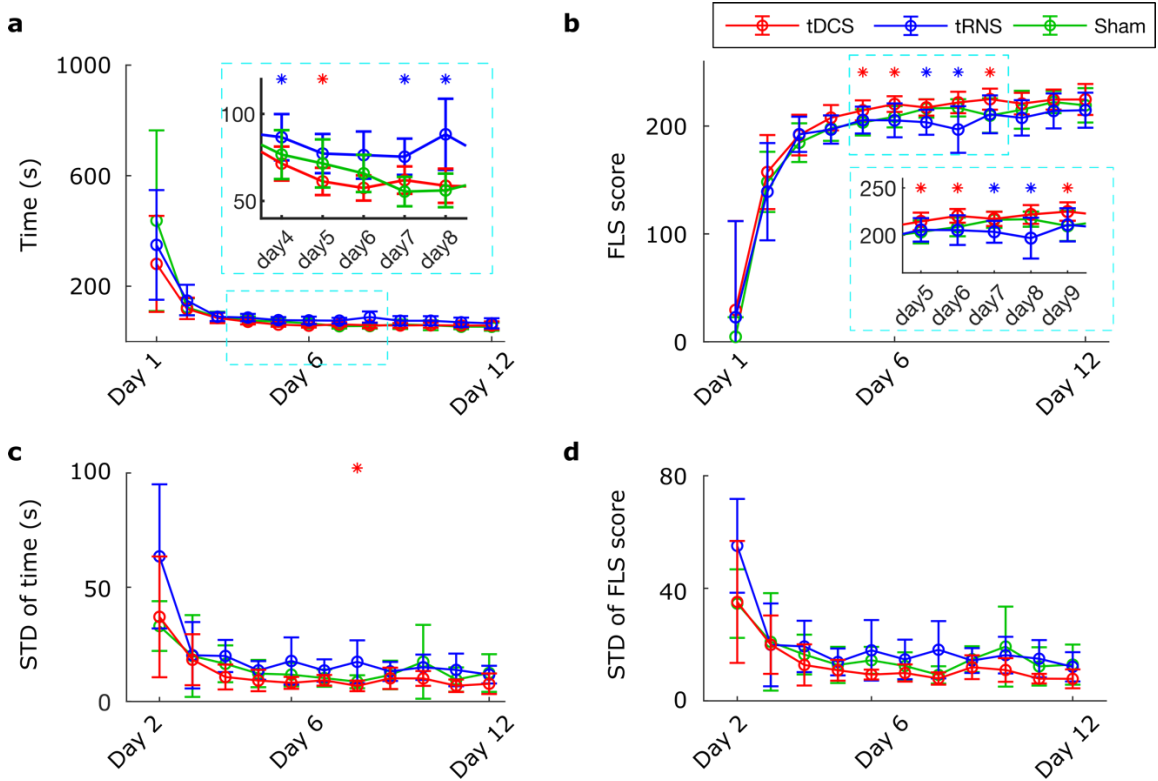

**Figure 3.** Bimanual motor task performance time and FLS score. (a) The performance time and (b) the FLS score for the 12 training days and the follow-up tasks are presented. (c-d) The standard deviation (STD) values of (c) time and (d) FLS scores are also displayed from day 2 to day 12. Green and blue boxes are the enlarged figures. The red lines represent the tDCS group, the blue lines represent the tRNS group, and the green lines represent the Sham group. The stars represent a significant difference ($\alpha = 0.05$) compared to the Sham group.

**Brain activation change during training**

The average of oxy-hemoglobin (HbO) changes are shown as spatial maps during the training day 2-6, and during day 7-12 in Fig. 4 (baseline activations on day 1 are not significantly different between three groups, see Fig. S4). For the Sham group, the right PFC and the left M1 region were activated during the initial learning stage (day 2-6), and the activation shifted to the M1 region with a higher activation level during the later learning stage (day 7-12). The brain activation patterns resulting from learning in the Sham group are consistent with findings from earlier studies (6, 13). The tDCS group showed a similar brain activation pattern as the Sham group, but with a slightly higher (left lateral M1 $p = 0.804$ and SMA $p = 0.254$) activation level in the initial learning stage (day 2-6). This difference in activation reached significance for days 7-12 (left lateral M1 $p = 0.038$, SMA $p < 0.001$). The most prominently activated area, *i.e.*, the left M1 region, was the cortical area under the tDCS anode. For the tRNS group, the left M1 region was lower activated compared to tDCS and Sham, with a depressed activation observed compared to the Sham group throughout the training (Significance observed during day 2-6: right PFC: $p < 0.000$; left lateral M1: $p = 0.017$; right middle M1: $p = 0.037$. Significance observed during day 7-12: left PFC: $p = 0.015$; right PFC: $p < 0.000$; right lateral M1: $p < 0.000$). At the late learning stage, tDCS and tRNS has unilateral and sham has more bilateral activation.

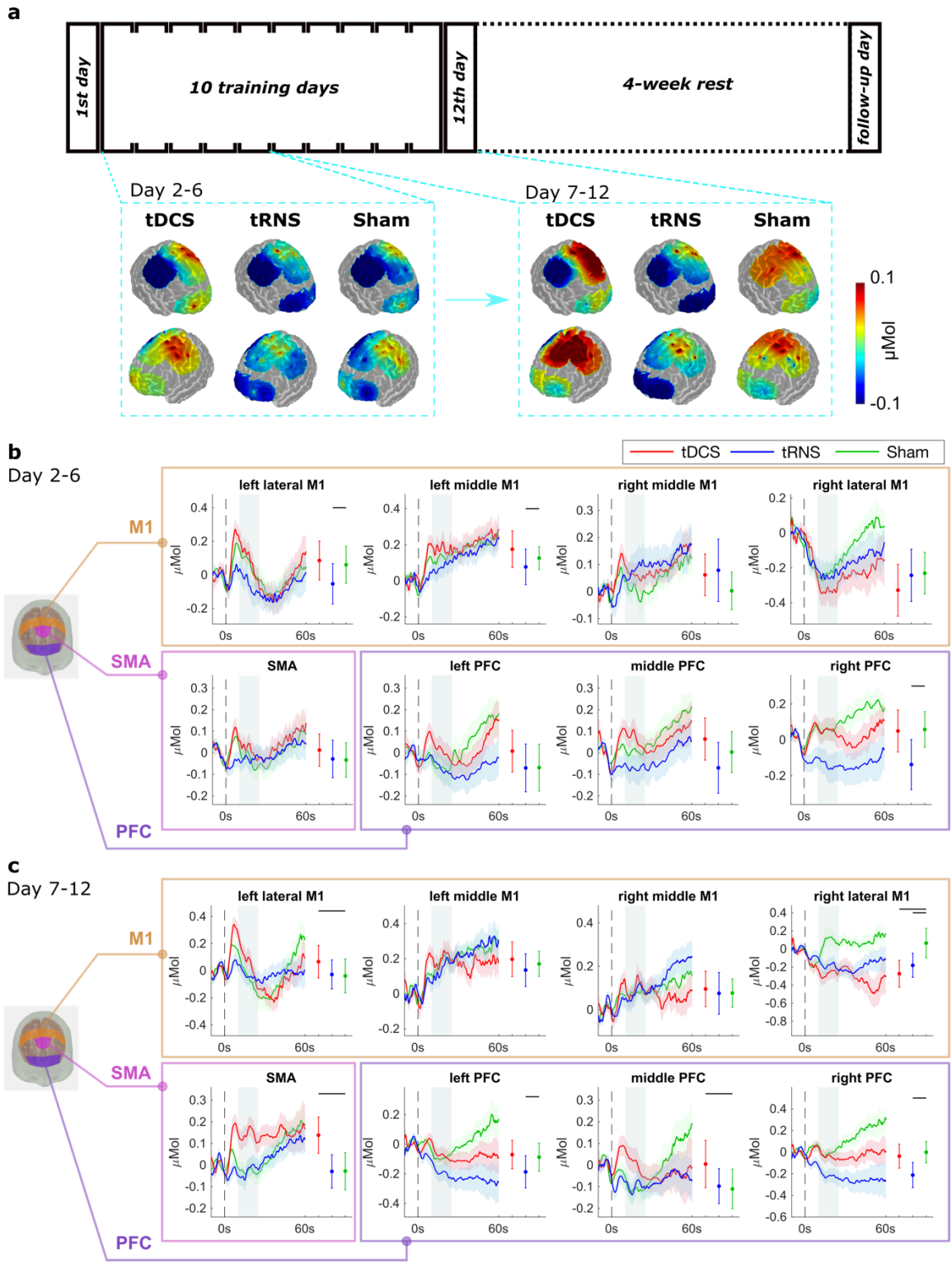

**Figure 4.** Brain functional activation during the training procedure. (a) The average of HbO changes is shown as spatial maps for PFC, M1, and SMA regions for training day 2-6, and day 7-12. (b) Grouped average time-series HRFs with respect to cortical regions on training day 2-6 and (c) day 7-12. The solid lines are mean

values, and the shaded areas are 95% confidence interval. The stimulus onset begins at zero seconds (dashed black line) indicating that the trial has started. Negative time indicates the baseline measurement used for calibration before each trial. The grey painted box (10-25s) is the time range selected to calculated the mean HbO values. The mean and 95% CI of 10-25s HRFs are plotted next to the time-series HRFs in error bar form. The black bar indicates $p < 0.05$.

**Skill retention**

The baseline performance (day 1), the post-test (day 12), and the retention (follow up performance) were analyzed, and the results are shown in Fig. 5b. For the performance error, the two-way mixed factorial ANOVA showed that there was a significant difference between stimulation types (tDCS, tRNS, and Sham; $F = 5.52; p = 0.005$), but not the effect of time points (day 1, day 12, and follow up visit; $F = 0.76; p = 0.469$), or the interaction of stim×time ($F = 1.14; p = 0.338$). For the performance time, the effect of stimulation ($F = 4.14; p = 0.018$), time point ($F = 168.2; p < 0.0001$), and the stim×time interaction ($F = 6.11; p = 0.0001$) were significant. For the FLS score, there was a significant difference between time points ($F = 333.56; p < 0.0001$), but not stimulation ($F = 0.06; p = 0.941$), or the interaction of stim×time ($F = 2.44; p = 0.050$).

The comparison of the performance between the three groups is shown in Fig. 5c. The tDCS and tRNS groups maintained significantly lower error levels than the Sham group ($p = 0.010$ and $p = 0.005$, respectively) after four weeks without training. This is consistent with the training performance error, *i.e.*, the tDCS and tRNS groups showed lower error than Sham. However, the tDCS group took longer to complete the follow-up task than Sham ($p = 0.036$). The FLS scores were not significantly different among the three groups. In summary, the difference in skill across the groups was retained only for performance error whilst the time and FLS score was indistinguishable.

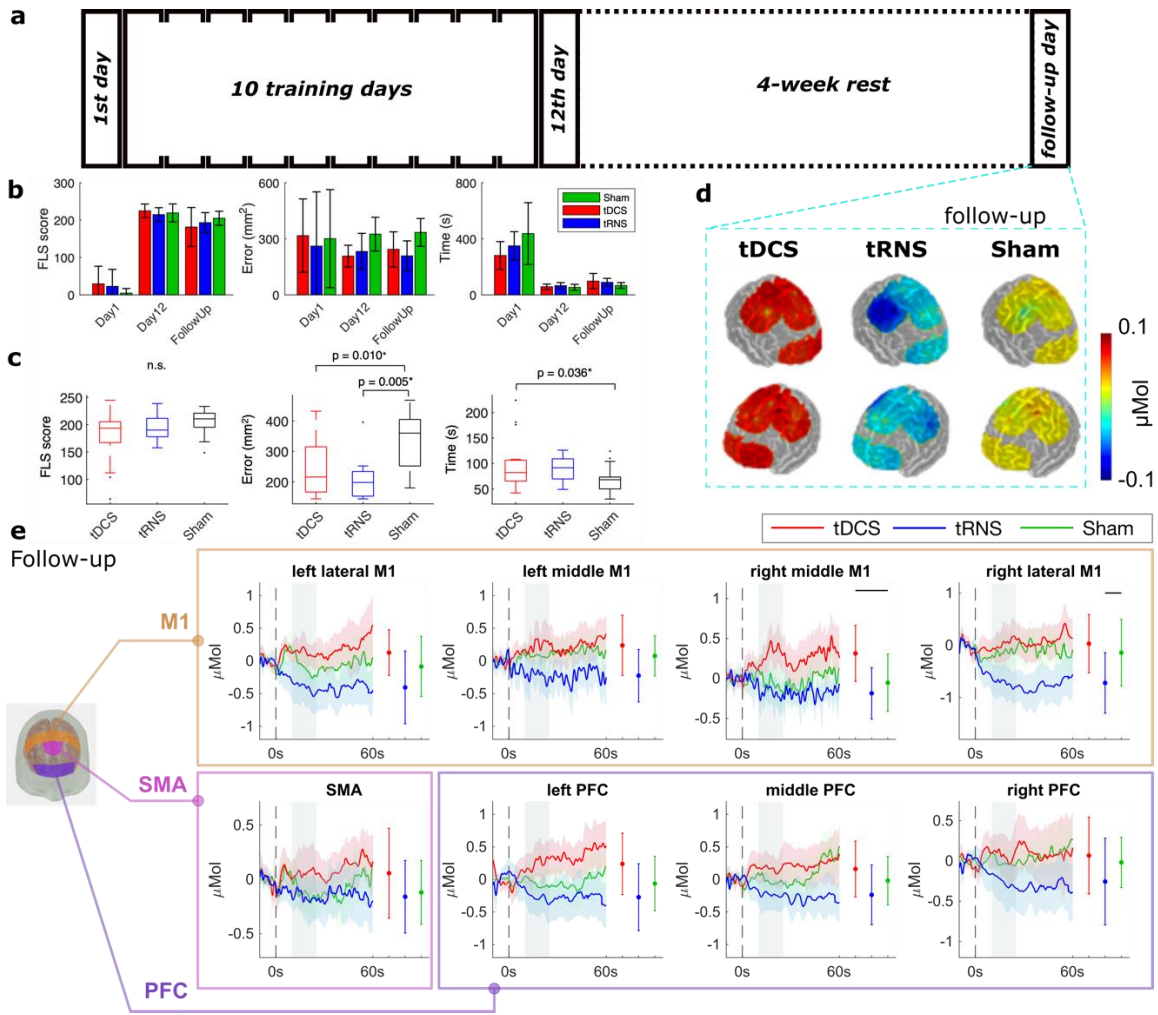

**Figure 5.** Bimanual motor task performance and functional brain activation for the follow-up tasks. (a) Experimental protocol. (b) The performance of the three groups on day 1, day 12, and the follow-up day visit in performance error, performance time, and the FLS score. (c) Bimanual motor task performance for the follow-up task. The stars represent a significant difference ($\alpha = 0.05$) compared to the Sham group. 'n.s.' represents not significant. (d) The average HbO changes are shown as spatial maps for PFC, M1, and SMA regions for follow-up tasks. (e) Grouped average time-series HRFs with respect to cortical regions on follow-up visit. The solid lines are mean values, and the shaded areas are 95% confidence interval. The stimulus onset begins at zero seconds (dashed black line) indicating that the trial has started. Negative time indicates the baseline measurement used for calibration before each trial. The grey painted box (10-25s) is the time range selected to calculated the mean HbO values. The mean and 95% CI of 10-25s HRFs are plotted next to the time-series HRFs in error bar form. The black bar indicates $p < 0.05$.

The comparisons among the three groups of their brain activation map on the retention task are shown in Fig. 5d with time-series results and significance test results in Fig. 5e. We observed a trend of increasing brain activation in the tDCS group and a decrease in the tRNS group, but the difference was not significant except tDCS in right medial M1 ($p = 0.014$), and tRNS in right lateral M1 ($p = 0.028$).

**Discussion**

In this study, we examined the effect of the M1 neurostimulation (tDCS and tRNS) on bimanual motor skill learning and retention. We also employed the fNIRS technique to acquire brain activation changes during task performance. We observed that tDCS lowered the performance error as the training proceeded. The tRNS group also performed at a lower error level, but the level remained the same throughout the training. The trial-to-trial standard deviation analysis demonstrated that the tDCS group stabilized performance variability in performance error, time, and FLS score, but the tRNS group only stabilized the performance error. The effect of tDCS and tRNS in enhancing performance accuracy was retained after a four-week break period. Accordingly, the M1 brain activation was enhanced by tDCS. Conversely, the overall brain activation was suppressed in the tRNS group. These fNIRS results show the different mechanisms of action between tRNS and tDCS, which has also been discussed in the literature (15, 16).

The effect that tDCS reduces performance errors observed in this study is significant since performance error is critical in surgical motor tasks. For example, surgery-related injuries were reported in 2.5 out of 1000 cases (17). The leading cause of maritime accidents was also reported to be human errors (49-85%) (18). From the learning curves shown in Figs. 2 and 3, the learners decreased their performance time, but error improvement occurred more slowly. Overall scores are weighted to value increased speed and miss the critical issue of reduced error. Once trained, the lowered error level was retained after four weeks (Fig 5). The two-way ANOVA analysis showed that the error did not change with the training itself, but did show a significant difference to the effect of neuromodulation. These results indicate the merit of the application of tDCS to increase performance accuracy in fine motor skill learning. The benefit, risks, and ethical dilemmas have been discussed in a recent review paper (11).

Existing literature supports our observations. M1 tDCS was observed to lower the hand path errors in an arm reaching task (19). In (20), a significantly lowered error was seen in motor sequence learning by M1 tDCS. It is worth mentioning that a study (21) also showed that tDCS was able to enhance the scores of the FLS pattern cutting task. However, their protocol only involved one-day training with eight repetitions of the pattern cutting task whereas, in our protocol, each participant practiced up to 10 trials per day from training day 2 to day 12, resulting in more than 100 repetitions. Therefore, our protocol involves various learning stages including acquisition, consolidation, and retention. Other previous studies have shown that learners could not reach proficiency in one day for laparoscopic skills (22). In our earlier analysis, even with the 12-day training on laparoscopic skills, some learners could not reach proficiency (12, 13). Here, we are the first to investigate the effect of tES during this 12-day training procedure. We did not see a significant difference in one training day, as in (21), comparing tDCS to the Sham group (the results are found in Fig. S7, which could be compared to (21)). The experimental settings are different in evaluating metrics (they assessed FLS score post-training, but we accessed the performance during the training) and tES settings (delivery timing: during the task in (21); here before the task).

Without any stimulation, in the Sham group, brain activation shifted to the bilateral M1 region with increased repetitions. Notably, fMRI studies have shown that the transition of short-term to long-term motor learning is featured by a shift of brain activation from anterior to more posterior brain regions (6). Training to perform an explicit sequence of finger movements over several weeks showed progressively increasing blood-oxygen-level-dependent (BOLD) activity in M1 (6, 23–28), interpreted as reflecting recruitment of additional M1 units into the local network that represents the acquired sequence of movements. In our previous study (13), we showed that skilled learners had more brain activation in M1 compared to unskilled learners, as well as expert surgeons compared to novices. With tDCS, we found more M1 activation under the anodal electrode, a phenomenon that has been reported in BOLD fMRI studies (6). The neurophysiology mechanism of tDCS has been discussed extensively in the literature (see a review paper (16)). To summarize it, the long-term effects of tDCS depend on neurotransmitters that bind to receptors, such as glutamatergic (NMDA) and gamma-Aminobutyric acid (GABA)-ergic receptors, as blocking these receptors resulted in suppression of the post-stimulation effects. In addition, glial cells and non-synaptic mechanisms may play a role in the afterward effects.

With tRNS, brain activation decreased. Though previous studies with tRNS are limited, one fMRI study confirms the decreased activation (29). In another study (15), the intracortical facilitation (ICF) was found to increase after tRNS over M1 using a paired-pulse paradigm, where increased ICF also indicates decreased HbO (30). In contrast to the after-effect of tDCS being NMDA- and GABA receptor-dependent, the tRNS effect might be related to repetitively open $Na^+$ channels, leaving the cell depolarized and unable to fire off another action potential (31). tRNS could also involve temporal summation of neural activity and interference with ongoing oscillations, which make the effect different from tDCS (32). The mechanism of tRNS could also be related to plasticity processes in high-frequency oscillations (80–200 Hz ripples) (33) or repeated subthreshold stimulations that prevent homeostasis of the system (32).

This study showed a successful combination of neuroimaging and neuromodulation. The multimodality of the two offered us valuable information that could not be derived with only one of them. The use of both neuroimaging and neuromodulation is an excellent way to investigate and understand neuromodulation (34). However, here, we only report the observation of the cortical activation changes via neuroimaging. Closer and more in-depth combination of the two could help boost the effect of the neuromodulation, such as a neurofeedback loop and personalized tES (34). Through these applications, the parameters of the neuromodulation could be regularized by the information from the neuroimaging technique. Motor performance could also be further optimized. This work contributes to this approach by offering the observation of the neural changes detected by neuroimaging under different neuromodulation conditions.

Another future direction is coupling neuromodulation with other techniques, such as EEG. fNIRS detects neurovascular changes in the cerebral cortex to reflect the brain activation level. However, brain activity includes other features, such as oscillatory, neuro-electrical, and neuro-chemical activities. The investigation in those areas could not be derived from fNIRS alone. Thus, introducing other measurements in the future could help understand the neuroscience mechanism of the neuromodulation, or the motor learning itself. Techniques other than neuroimaging are also beneficial. For example, the video data of the task execution or kinematic data of the tools also offer insights into motor learning.

In summary, we demonstrate that tDCS facilitates surgical bimanual motor skill learning by lowering the performance error. We further show enhanced anodal excitation of the M1 cortex with tDCS compared to Sham. While existing metrics of task performance reward increased execution speed, our work sheds light on the importance of reducing errors, and the positive role of tDCS in achieving that outcome.

## Materials and Methods

### Participant recruitment

This study was approved by the Institutional Review Boards of the University at Buffalo and Rensselaer Polytechnic Institute. Potential participants were recruited from the University at Buffalo's current medical student population through campus emails, social media, flyers, and word of mouth communication. The recruitment procedure is illustrated in Fig. S1. We pre-screened 30 respondents through a pre-screening questionnaire (35). Four of them were excluded due to safety concerns or hairstyles that might prevent electrode or optode contact with the scalp. Two dropped out before consenting to participate, and three dropped out during the study due to personal reasons, leaving twenty-one medical students who completed the training protocol. All participants were novices to the bimanual task, as they had no experience with laparoscopic tools, FLS training, or any similar surgical simulation training software. The power analysis could be found in supplemental material.

**Power analysis**

From our previous study (13), 12 days of FLS pattern cutting task training showed clear learning curves for both FLS score and fNIRS metrics. There was no previous study data that could support the power analysis to estimate the number of participants needed. Based on our previous study (13), an effect size was selected as Cohen's f of 1.26. For the ANOVA test power analysis with a 95% confidence interval and a minimum power of 0.80, it was determined that a minimum of 15 as the total sample size, calculated using the statistical software package G*Power. Among the 21 medical students recruited in this study, four subjects did not pass the CUSUM exam and were excluded from the data analysis (explained in the section 'Motor task and task performance metrics'). The subjects were divided into three groups randomly at the beginning of the study: tDCS, tRNS, and Sham. The resulted sample population was distributed across three different neuromodulation types: tDCS (n = 5, mean age 24 ± 3), tRNS (n = 5, mean age 23 ± 1) and Sham (n = 7, mean age 23 ± 1) (demographics in Table S1).

**Experimental design**

The participants underwent 12 visits on 12 consecutive days and one visit as a follow-up visit four weeks later. On the first day, demographic information, including age and handedness, were collected. Participants were made familiar with the experimental apparatus, including the cap, optodes, and electrodes. The impedance of tES and the signal quality of fNIRS were checked to ensure they were within an acceptable range. All participants were instructed on how to perform the task through a standardized video tutorial (https://youtu.be/mUBZoSO3KA8) and verbal instructions. The bimanual task was the pattern cutting task selected from the FLS program (www.flsprogram.org) (13). After the training, participants performed a single trial of the pattern cutting task to show that they fully understood the task and as a measure of baseline performance. From day 2 to day 12, the subjects underwent 10 minutes of the stimulation (tDCS/tRNS/Sham according to their group assignment) and then 30 minutes of practice on the task, while being imaged with fNIRS. Four weeks after the completion of the training, subjects returned for a follow-up visit, during which they performed the same FLS pattern cutting task three times, to measure their skill retention. Participants completed a safety questionnaire (35) before and after each neuromodulation session (Fig. S5). A photo of a participant during the experiment is shown in Fig. S1b.

**Motor task and task performance metrics**

The participants practiced a bimanual motor task, pattern cutting, selected from the FLS program (www.flsprogram.org) which is a pre-requisite for board certification in general and obstetrics and gynecology surgery. The bimanual motor task was performed using an FLS simulator (Laprascopic skills trainer, Limbs & Things, UK). Standard FLS-certified laparoscopic tools were used to cut a marked piece of gauze as quickly and as accurately as possible. The participants were instructed to avoid unnecessary movements of their body or facial muscles, and to refrain from speaking, to prevent motion artifacts in the fNIRS signals. The cap holding the fibers on the participant, as well as the wires, did not hinder the participant's movement during the task.

The three performance quantification metrics were time, errors, and the FLS score. Time was captured starting from the moment when the tools touched the gauze and ending at the moment of completion. The error was counted as the area between the marked circle on the gauge and the actual cut. The FLS scores were determined using the standardized FLS scoring metric formulation for the pattern cutting task based on the time and error. This formulation is intellectual property–protected and was obtained with a nondisclosure agreement with the FLS Committee, and hence, its details cannot be reported in this paper.

After the training, we plotted a CUSUM learning curve for each participant. In CUSUM analysis, positive or negative increments are added to a cumulative score according to the failure or success of the successive trial (36). The pass/fail threshold was set at the average score they received on the first five trials (FLS score of 33), a method derived from (37). This value is lower than 44 in the non-competent group in (38) and 39 in the novice group in (39). However, considering that in this study, we only recruited medical students without any experience, less experienced than the cohort of medical students and junior surgical residents in (38) and surgeons of PGY1-4 in (39), a lower threshold is reasonable. To calculate the CUSUM score, if a trial was a "pass," the respective CUSUM score is subtracted by 0.07; the CUSUM score was incremented by 0.93 in case of a "fail". Four subjects did not pass the desired acceptable failure rate of 0.05 and were excluded from the data analysis. A graphical representation of the CUSUM curve for each participant is in Fig. S1b. The brain activation pattern of the excluded subjects was in Fig. S1c. The pattern is similar to the follow-up task in Fig. 5d, in the sense that the brain cortex was over-activated and the activation was spread out through the cortex regions.

**tES hardware and settings**

The tES stimulation was delivered by a commercial device (StarStim, Neuroelectronics, Spain). One electrode was placed over the C3 location, according to the international 10-20 system, aiming to simulate the left primary motor cortex (M1) region. The return electrode was placed at Pf2 over the right PFC region. The electrode was based on a sintered Ag/AgCL pellet with a 12 mm diameter, with a total electrode area of 1 cm2. Gel (Parker Laboratories, Inc) was applied on the electrode to decrease the impedance and improve the signal quality. A trained researcher carefully adjusted the contact between the electrode and the scalp until the impedance was lower than 15 $k\Omega$. During the stimulation protocol, the impedance was monitored every second. At any time, if the impedance exceeded 20 $k\Omega$ in any stimulation electrode, the stimulation protocol was aborted to protect the subject from the high voltages generated as an embedded function in StarStim. The stimulation lasted 10 minutes. tDCS was delivered at 1mA. The current was ramped up for 30 seconds to 1mA and down to zero current at the beginning and the end of the stimulation. tRNS was performed at 1 mA, 0.1-650 Hz. Sham stimulation was set at zero current with the same ramp parameters as tDCS, imitating the same sensation to blind the subjects to the stimulation type (see Fig. S2d).

**NIRS Hardware and equipment**

We used a commercially available, continuous-wave near-infrared spectrometer, which delivered infrared light at 760 and 850 nm (NIRScout, NIRx, Berlin, Germany). The system used 8 illumination fibers coupled to 19 long-distance detectors and 8 short-distance detectors. The combinations between the illumination fibers and the detectors resulted in 28 measurement channels. A schematic of the geometric arrangement of probes is in Fig. S2a. The long-distance channels captured all measurements within a 30- to 40-mm distance between the source and the detector. The short-distance channels captured measures within an ~8-mm distance between the source and the detector. The short-distance channels were limited to probing the superficial tissue layers, such as skin, bone, dura, and pial surfaces, whereas the long-distance channels penetrated both superficial layers and cortical surface. The probe design was derived from our previous study (13), which was determined using Monte Carlo simulations and was characterized to have high sensitivity to functional changes in the PFC, M1, and SMA. We further added one channel to each of the SMA, the left, and the right lateral M1 regions to increase data measurement in those regions of particular interests. The optical probes were positioned carefully on the participant to avoid hair between the source/detector and scalp.

**Optode positioning**

The method to ensure the optode position placement in this study was two-fold. First, we used a standard electroencephalography cap (EASYCAP: www.easycap.de) to hold the optodes (13). The cap had marked anatomical landmarks for placement on the scalp. The cap was carefully placed on the scalp by aligning the landmarks on the head. Second, we used a magnetic tracking device (trakSTAR, Ascension Tech Corp. Canada) to measure the positions of the optodes in 3D space. After we put the cap with optodes and electrodes on a participant, a 6 DOF sensor (diameter = 2mm) was placed on the positions of the optodes and electrodes to measure the 3D positions. The measured data were analyzed offline using the software package AtlasViewer to calculate the inter-trial standard deviation of the position for each optode to ensure that the optode was in the intended location. The mean positions are shown in Fig. S6 and Table S2.

**NIRS signal processing**

Data processing was completed using the open-source software HomER2 (40), which is implemented in MATLAB (MathWorks). First, channels with light intensity greater than $10 \, \mu V$, which indicates environmental light contaminated the signals, or smaller than $0.01 \, \mu V$, which indicates the light was blocked by hair, were excluded. The channels with signal-to-noise ratio of light intensity greater than 3 were excluded as well due to poor signal quality. The remaining raw optical signals (intensity at 760 and 850 nm) were converted into optical density. Motion artifacts and systemic physiological interference were corrected using a third-order Butterworth low-pass filter with a cutoff frequency of 0.5Hz. The filtered optical density data were used to derive the concentration changes of oxyhemoglobin and deoxyhemoglobin by Beer-Lambert Law (40). The short-distance channels were regressed from the long-distance channels to remove any interference originating from superficial layers. This was achieved by using a consecutive sequence of Gaussian basis functions ($stdev = 1\mu Mol$; $step = 1s$) via ordinary least squares to regress scalp and dura activation data collected from the short separation fibers, to create the hemodynamic response function (HRF) (40). Then, the time series data for each of the 28 channels were group-averaged into eight distinct regions of interest as follows, according to the anatomical structures: left PFC (source 1, detectors 1 and 2), medial PFC (source 2, detectors 2 and 3), right PFC (source 3, detectors 3 and 4), left lateral M1 (source 4, detectors 5 to 8 and 17), LMM1 (source 5, detectors 8 to 10), right medial M1 (source 6, detectors 9 to 12), right lateral M1 (source 7, detectors 11 to 14 and 18), and finally, SMA (source 8, detectors 9, 15, 16 and 19).

The time range used from the fNIRS time series to calculate the brain activation level was determined by Linear Discriminate Analysis (LDA). The LDA was used to classify three groups based on the averaged oxy-hemoglobin (HbO) value within time ranges, with an exhaustive grid search on the starting time point and ending time point, with an increment of 5s. Then we selected the time range of 10s – 20s, which yielded the largest LDA accuracy, as shown in Fig. S3.

**Statistical methods**

The mean of performance error, performance time, and the FLS score was compared between the three groups (tDCS, tRNS, and Sham) using a series of one-way ANOVAs, on each training day and for the retention task. Two-way ANOVA was adopted when analyzing the effect of stimulation type and the training time points on the performance. The mean brain activation was also compared between groups by one-way ANOVA for each of the eight brain areas. Post hoc pairwise comparison was conducted using Tukey's honestly significant difference procedure. An alpha level of 0.05 was set as the minimum required to reject the null hypothesis. Descriptive and inferential statistics were performed in MATLAB. All error bar plots display mean values along with a 95% confidence interval of the mean values. Ordinary least squares were used to evaluate the change of errors along the training days, with the training day as the independent variable, and the performance error as the dependent variable.

**Data and code availability:**
The data related to this study is available upon request to the corresponding authors; The code has been uploaded in https://github.com/YuanyuanGao216/NeuromodulationStudy.

**Acknowledgments**

We thank the funding provided by NIH/National Institute of Biomedical Imaging and Bioengineering grants 2R01EB005807, 5R01EB010037, 1R01EB009362, 1R01EB014305, and R01EB019443.

**Supplemental Material**

Fig. S1. Participant recruitment.

Fig. S2. fNIRS and tES setups.

Fig. S3. Select the time range along the fNIRS time series data.

Fig. S4. Brain functional activation for training day 1.

Fig. S5. The safety questionnaire data.

Fig. S6. The averaged optode position for fNIRS.

Fig. S7. The FLS score for each trial on day 2.

Table S1. Participant demographics.

Table S2. Optode 3D positions and spatial standard deviations (mm).

35. H. Thair, A. L. Holloway, R. Newport, A. D. Smith, Transcranial direct current stimulation (tDCS): A beginner's guide for design and implementation. *Front. Neurosci.* **11**, 641 (2017).

36. I. G. Kestin, A statistical approach to measuring the competence of anaesthetic trainees at practical procedures. *Br. J. Anaesth.* **75**, 805–809 (1995).

37. A. M. Linsk, *et al.*, Validation of the VBLaST pattern cutting task: A learning curve study. *Surg. Endosc. Other Interv. Tech.* **32**, 1990–2002 (2018).

38. S. A. Fraser, *et al.*, Evaluating laparoscopic skills, setting the pass/fail score for the MISTELS system. *Surg. Endosc. Other Interv. Tech.* **17**, 964–967 (2003).

39. A. Chellali, *et al.*, Preliminary evaluation of the pattern cutting and the ligating loop virtual laparoscopic trainers. *Surg. Endosc.* **29**, 815–821 (2015).

40. T. J. Huppert, S. G. Diamond, M. A. Franceschini, D. A. Boas, HomER: A review of time-series analysis methods for near-infrared spectroscopy of the brain. *Appl. Opt.* **48**, D280–D298 (2009).
19